\documentclass[twocolumn,aps,superscriptaddress]{revtex4-1}
\usepackage{graphicx,amsmath,amsfonts}
\begin{document}

\title{Microscopic toy model for magnetoelectric effect in polar Fe$_2$Mo$_3$O$_8$}

\author{I.~V. Solovyev}
\email{SOLOVYEV.Igor@nims.go.jp}
\affiliation{International Center for Materials Nanoarchitectonics,
National Institute for Materials Science, 1-1 Namiki, Tsukuba,
Ibaraki 305-0044, Japan}
\affiliation{Department of Theoretical Physics and Applied Mathematics, Ural Federal University,
Mira str. 19, 620002 Ekaterinburg, Russia}

\author{S.~V. Streltsov}
\affiliation{Department of Theoretical Physics and Applied Mathematics, Ural Federal University,
Mira str. 19, 620002 Ekaterinburg, Russia}
\affiliation{Institute of Metal Physics, S. Kovalevskaya str. 18, 620108 Ekaterinburg, Russia}
\date{\today}

\date{\today}
\begin{abstract}
The kamiokite, Fe$_2$Mo$_3$O$_8$, is regarded as a promising material exhibiting giant magnetoelectric (ME) effect at the relatively high temperature $T$. Here, we explore this phenomenon on the basis of first-principles electronic structure calculations. For this purpose we construct a realistic model describing the behavior of magnetic Fe $3d$ electrons and further map it onto the isotropic spin model. Our analysis suggests two possible scenaria for Fe$_2$Mo$_3$O$_8$. The first one is based on the homogeneous charge distribution of the Fe$^{2+}$ ions amongst tetrahedral ($t$) and octahedral ($o$) sites, which tends to low the crystallographic P6$_3$mc symmetry through the formation of an orbitally ordered state. Nevertheless, the effect of the orbital ordering on interatomic exchange interactions does not seem to be strong, so that the magnetic properties can be described reasonably well by averaged interactions obeying the P6$_3$mc symmetry. The second scenario, which is supported by obtained parameters of on-site Coulomb repulsion and respects the P6$_3$mc symmetry, implies the charge disproportionation involving somewhat exotic $1+$ ionization state of the $t$-Fe sites (and $3+$ state of the $o$-Fe sites). Somewhat surprisingly, these scenarios are practically indistinguishable from the viewpoint of exchange interactions, which are practically identical in these two cases. However, the spin-dependent properties of the electric polarization are expected to be different due to the strong difference in the polarity of the Fe$^{2+}$-Fe$^{2+}$ and Fe$^{1+}$-Fe$^{3+}$ bonds. Our analysis uncovers the basic aspects of the ME effect in Fe$_2$Mo$_3$O$_8$. Nevertheless, the quantitative description should involve other ingredients, apparently related to the lattice and orbitals degrees of freedom.
\end{abstract}

%\keywords{}
\maketitle

\section{\label{sec:Intro} Introduction}
\par Materials with the general formula Me$_1$Me$_2$Mo$_3$O$_8$, where Me$_1$ and Me$_2$ are alkali, alkali earth, transition or post-transition metal ions distributed amongst tetrahedral and octahedral positions, are extremely interesting not only for the fundamental science, but also for different applications. Various intriguing phenomena such as realization of the spin-liquid phase~\cite{Haraguchi2015}, giant optical diode effect~\cite{Yu2018}, valence-bond condensation~\cite{Sheckelton2012}, and magnetoelectricity~\cite{Wang_SciR,Kurumaji_PRX} were found in this group of materials.  Such a variety is ultimately related to three aspects of the crystal structure of Me$_1$Me$_2$Mo$_3$O$_8$. First, it is polar, which is important for the magnetoelectric effect. Second, the Me$_1$ and Me$_2$ sites can easily accommodate all kind of ions, starting from the simple alkali ones, and ending by transition or even post-transition metal elements. As a result, by changing Me$_1$ and Me$_2$ one may vary the valency of Mo ions. Furthermore, the Mo ions form isolated trimers (the third important aspect), which makes these materials interesting testbed also for studying of the cluster-Mott physics~\cite{Chen2018,Streltsov2017}.

\par Fe$_2$Mo$_3$O$_8$ (the kamiokite~\cite{Sasaki1985}) is one of such materials, whose properties were under intensive investigation during last years. The Fe ions in Fe$_2$Mo$_3$O$_8$ occupy both tetrahedral ($t$-Fe) and octahedral ($o$-Fe) positions. Furthermore, the FeO$_4$ tetrahedra are distorted and this distortion points in the same ($z$) direction~\cite{Fe2Mo3O8str}. Thus, the material is polar and this property is manifested in the nonreciprocal high-temperature optical diode effect, which was observed in Zn doped Fe$_2$Mo$_3$O$_8$, where the intensity of light transmitted in one of the directions was hundred times smaller than in the opposite one~\cite{Yu2018}.

\par Another interesting aspect of Fe$_2$Mo$_3$O$_8$ is the magnetoelectric properties -- the interplay of the electric polarization and magnetism. Due to the trimerization, the Mo$^{4+}$ ions appear to be nonmagnetic. However, the Fe ions have local magnetic moments, which order antiferromagnetically below $T_{\rm N} \sim $ 60~K~\cite{Czeskleba1972}. The antiferromagnetic (AFM) transition is accompanied by the giant ($\sim 0.3$ $\mu$C/cm$^2$) jump of the electric polarization~\cite{Wang_SciR}. Furthermore, the AFM order appears to be fragile and can be easily switched to the ferrimagnetic (FRM) one by the external magnetic field and/or the Zn doping~\cite{Bertrand1975,Kurumaji_PRX,Wang_SciR}. This AFM-FRM transition is again accompanied by the jump of electric polarization being of the order of $-0.1$ $\mu$C/cm$^2$~\cite{Wang_SciR,Kurumaji_PRX}. These examples clearly show that the electric polarization in Fe$_2$Mo$_3$O$_8$ depends on the magnetic order and can be manipulated by changing the magnetic order. Another interesting manifestation of the magnetoelectric coupling in Fe$_2$Mo$_3$O$_8$ is the observation of electromagnons~\cite{Kurumaji2017}.

\par Although the electronic structure of Fe$_2$Mo$_3$O$_8$ and related (Fe,Zn)$_2$Mo$_3$O$_8$ compound was thoroughly investigated both experimentally and theoretically~\cite{Kurumaji2017-2,Yu2018,Stanislavchuk2019}, details of the exchange coupling responsible for the AFM-FRM transition remain mostly unexplored. Furthermore, there is no clear consensus on the microscopic origin of giant magnetoelectric effect observed in Fe$_2$Mo$_3$O$_8$. Originally, it was attributed to the magnetostriction, which manifests itself in different atomic displacements in different magnetic states~\cite{Wang_SciR}. Nevertheless, an alternative point of view based on the Dzyaloshinkii-Moriya mechanism was proposed recently in Ref.~\cite{Li2017}.

\par In this paper we study magnetic properties and magnetoelectric effect in Fe$_2$Mo$_3$O$_8$ using first-principles electronic structure calculations. After brief discussion of the electronic structure of Fe$_2$Mo$_3$O$_8$ in Sec.~\ref{sec:structure}, in Sec.~\ref{sec:model} we will discuss the construction the simple but realistic model describing the behavior of magnetic Fe $3d$ electrons. It can be regarded as the microscopic toy model for Fe$_2$Mo$_3$O$_8$, which included explicitly neither O $2p$ nor Mo $4d$ states. The main advantage of this model is its transparency, which can be regarded as the possible alternative to the local density approximation (LDA) $+$$U$ methods~\cite{LDAU}, which are formulated in the complete basis set of states, but suffer from uncertainty with the choice of parameters specifying the subspace of correlated electrons~\cite{PRB98}, and in this sense is less transparent.  Then, the effective $3d$ model is further mapped onto the isotropic spin model (Secs.~\ref{sec:solution}, \ref{sec:J}, and \ref{sec:P}), which is analyzed in terms of molecular-field approximation (MFA, Sec.~\ref{sec:discussion}).

\par Our analysis suggests two possible scenarios for Fe$_2$Mo$_3$O$_8$. The first one is based on the homogeneous charge distribution amongst tetrahedral ($t$) and octahedral ($o$) Fe sites ($d_{t}^{6}d_{o}^{6}$, denoting the formal number of Fe $3d$ electrons at these two types of sites), which tends to low the crystallographic P6$_3$mc symmetry through the formation of an orbitally ordered state. Nevertheless, the effect of the orbital ordering on the interatomic exchange interactions does not seem to be crucial and the magnetic properties can still be approximately described by averaged interactions obeying the P6$_3$mc symmetry. The second scenario implies the charge disproportionation, $d_{t}^{7}d_{o}^{5}$, involving somewhat exotic Fe$^{1+}$ ionization state. Nevertheless, it is supported by obtained parameters of on-site Coulomb interactions, which are more ``repulsive'' at the $o$-Fe sites, reflecting details of the electronic structure. Furthermore, it respects the crystallographic P6$_3$mc symmetry. Somewhat surprisingly, these two scenarios are practically indistinguishable from magnetic point of view as they produce very similar sets of parameters of interatomic exchange interactions. However, the spin-dependent properties of the electric polarization are rather different, due to the strong difference in the polarity of the Fe$^{2+}$-Fe$^{2+}$ and Fe$^{1+}$-Fe$^{3+}$ bonds, realized in the case of $d_{t}^{6}d_{o}^{6}$ and $d_{t}^{7}d_{o}^{5}$, respectively. The MFA uncovers the basic aspects of the ME effect in Fe$_2$Mo$_3$O$_8$, related to the emergence of net magnetization at finite temperature $T$, which can be controlled by the magnetic field, thus inducing the antiferromagnetic-to-ferrimagnetic phase transition.

\par Finally, the brief summary of our work will be given in Sec.~\ref{sec:conc}. According to our analysis, the magnitude of the magnetoelectric effect in Fe$_2$Mo$_3$O$_8$ can be understood by considering the isotropic electronic contributions to the electric polarization for the fixed crystal structure, though the quantitative description of the temperature dependence of both magnetization and polarization should probably include the lattice effects~\cite{Wang_SciR}.

\section{\label{sec:Method} Method}
\subsection{\label{sec:structure} Electronic structure in LDA}
\par
\begin{figure*}[t!]
\begin{center}
\includegraphics[width=0.8\textwidth]{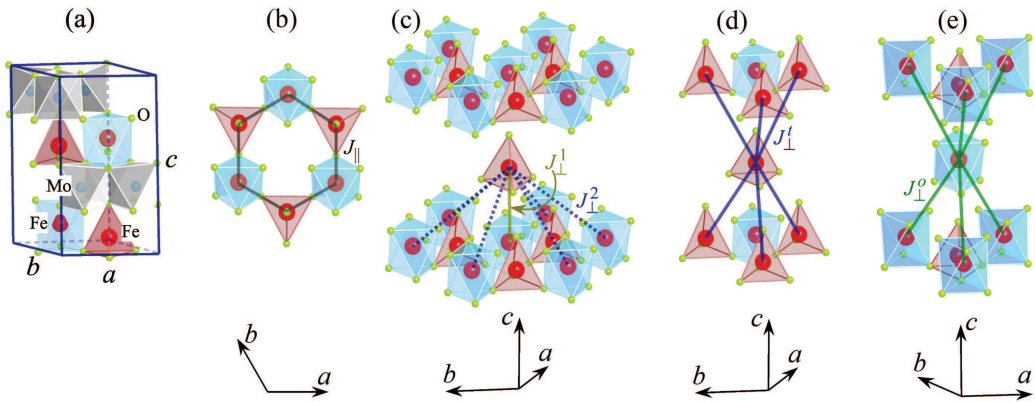}
\end{center}
\caption{
Fragments of the crystal structure of Fe$_2$Mo$_3$O$_8$ with the notations of main exchange interactions: (a) alternation of honeycomb layers formed by FeO$_{4}$ tetrahedra FeO$_{6}$ octahedra and trimerized kagome layers of MoO$_6$ octahedra in the unit cell of Fe$_2$Mo$_3$O$_8$; (b) nearest-neighbor interactions in the honeycomb layers; (c) interlayer interactions between tetrahedral and octahedral Fe sites located in the first ($J_{\perp}^{1}$) and second ($J_{\perp}^{2}$) coordination spheres; (d) and (e) interlayer interactions between tetrahedral and octahedral Fe sites, respectively. Fe, Mo, and O atoms are denoted by large, medium, and small spheres respectively. }
\label{fig.str}
\end{figure*}

 The crystal structure of Fe$_2$Mo$_3$O$_8$ (the space group P6$_3$mc, No. 186) consists of the honeycomb-like layers formed by the corner-sharing FeO$_4$ tetrahedra and FeO$_6$ octahedra, which are separated by trimerized kagome-like layers of the MoO$_6$ octahedra, as explained in Fig.~\ref{fig.str}.
\noindent

\par
\begin{figure}[h!]
\begin{center}
\includegraphics[width=0.47\textwidth]{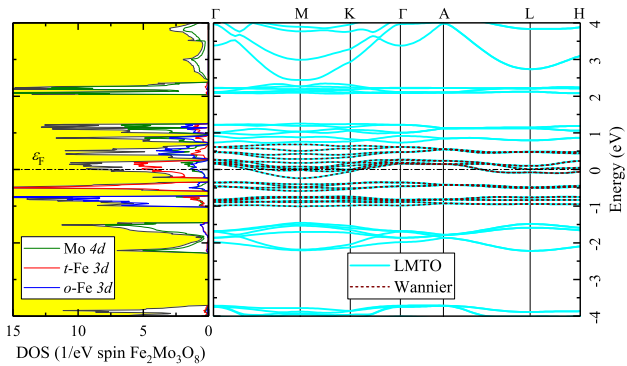}
\end{center}
\caption{
(Left panel) Total and partial densities of states in the local density approximation. (Right panel) Corresponding band structure calculated in the full LMTO basis (solid curved) and in the Wannier basis for the Fe $3d$ bands. The Fermi level is at zero energy (shown by dot-dashed line). Notations of the high-symmetry points of the Brillouin zone are taken from Ref.~\cite{BradlayCracknell}.}
\label{fig.elstr}
\end{figure}

 We use the linear muffin-tin orbital (LMTO) method~\cite{LMTO1,LMTO2} and the experimental structure parameters reported in Ref.~\cite{Fe2Mo3O8str}. The practical aspects of calculations (including the choice of atomic sphere, etc.) can be found in Ref.~\cite{LMTO_details}. The corresponding band structure in LDA is shown in Fig.~\ref{fig.elstr}.
\noindent
\noindent Some test calculations have been also performed using the full potential Wien2k method~\cite{Wien2k}, which reveals a good agreement with the LMTO results, as discussed in Supplemented Materials~\cite{SM}.

\par
\begin{figure}[b!]
\begin{center}
\includegraphics[width=0.47\textwidth]{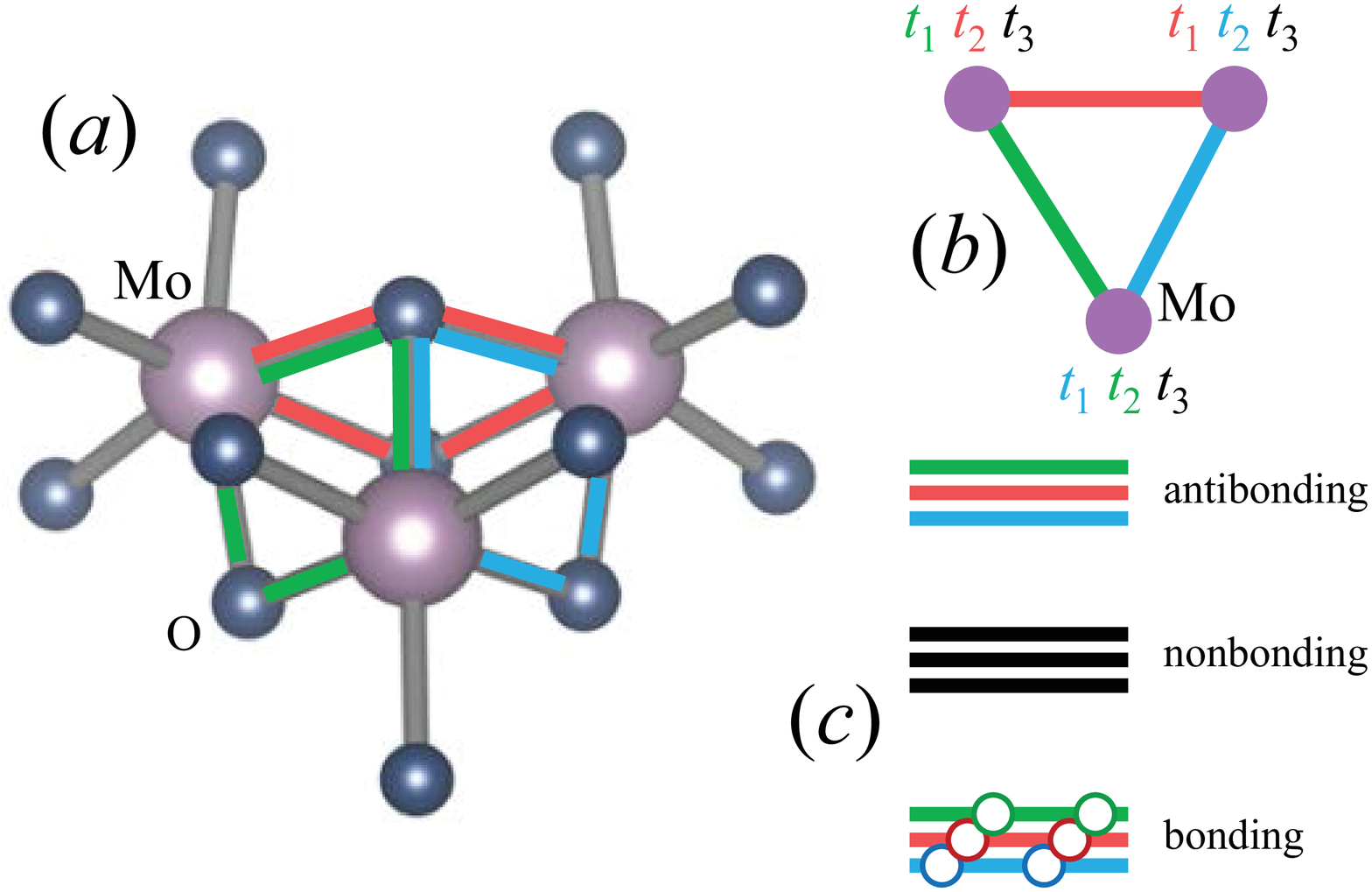}
\end{center}
\caption{
(a) The Mo$_3$O$_{13}$ cluster with the notations of Mo-O-Mo paths mediating the hybridization between $t_{2g}$ orbitals in each of the Mo-Mo bond. (b) Schematic view on the hybridization in the Mo$_{3}$ trimer: each Mo site donates one $t_{2g}$ orbital for the hybridization in each of the Mo-Mo bonds, resulting in the formation of bonding and antibonding molecular states. These orbitals are denoted as $t_1$ and $t_2$ and shown by the color of the bond in which they operate. The third $t_{2g}$ orbital is nonbonding and denoted as $t_{3}$. (c) Schematic view of the bonding-nonbonding-antibonding splitting in the Mo$_{3}$ trimer resulting in the nonmagnetic state, where six $4d$ electrons of Mo$_{3}$ reside at the bonding molecular orbitals. The molecular levels are shown by the same color as forming them atomic orbitals.}
\label{fig.Mo}
\end{figure}

 Owing to the trimerization of Mo kagome-like layers~\cite{Wang_SciR}, the Mo $4d$ states form well separated groups of $t_{2g}$ bands each of which corresponds to the particular type of molecular orbitals. This can be understood as follows. The formal configuration of octahedrally coordinated Mo$^{4+}$ ions is $t_{2g}^{2}$. If intersite hybridization is larger than the crystal field, as in the Mo$_3$ trimer, two $t_{2g}$ orbitals ($t_1$ and $t_2$ in Fig.~\ref{fig.Mo}) at each Mo site can be chosen so to form the maximal overlap with either $t_1$ or $t_2$ orbitals of the neighboring Mo site, where each orbital participates in the hybridization in only one Mo-Mo bond, as schematically illustrated in Fig.~\ref{fig.Mo}(b).
\noindent
\noindent In reality, such hybridization can occur via the Mo-O-Mo paths of the edge sharing MoO$_6$ octahedra, as shown in Fig.~\ref{fig.Mo}(a) or directly, as shown in Fig.~\ref{fig.Mo}(b). Therefore, in each of the Mo-Mo bonds, the atomic $t_1$ and $t_2$ orbitals will form bonding and antibonding molecular states, which are schematically shown in Fig.~\ref{fig.Mo}(c). Then, the third $t_{2g}$ orbital ($t_3$ in Fig.~\ref{fig.Mo}) will be nonbonding. In solids, these molecular levels will form bands, which can be still classified as bonding (at around $-1.8$ eV in Fig.~\ref{fig.elstr}), nonbonding (at around $1$ eV), and antibonding (at around $2.1$ eV). Since the bonding-nonbonding-antibonding splitting is much larger than the Hund's coupling $J$ (typically, about $0.4$ eV for Mo), the system will remain nonmagnetic with six $t_{2g}$ electrons of the Mo$_3$ trimer residing at the bonding orbitals.

\par The magnetic Fe $3d$ bands, which are located near the Fermi level, in the energy interval of about $[-1.0,0.8]$ eV, are sandwiched between bonding and nonbonding Mo bands. The Fe $3d$ and Mo $4d$ bands are separated from each other by a finite energy gap, which makes straightforward the construction of the effective model for the Fe $3d$ bands. Furthermore, there are two groups of the Fe $3d$ bands: the $t$-Fe one, which is formed mainly by the tetrahedral sites and located closer to the Fermi level, and the $o$-Fe bands, formed by the octahedral sites, which are split and located away from the Fermi level.

\subsection{\label{sec:model} Effective model for the Fe $3d$ bands}
\par The effective Hubbard-type model for the magnetic Fe $3d$ bands,
\noindent
\begin{widetext}
\begin{equation}
\hat{\cal{H}}  =  \sum_{ij} \sum_{\sigma \sigma'} \sum_{ab}
t^{ij}_{ab}\delta_{\sigma \sigma'}
\hat{c}^\dagger_{i a \sigma}
\hat{c}^{\phantom{\dagger}}_{j b \sigma'} +
  \frac{1}{2}
\sum_{i}  \sum_{\sigma \sigma'} \sum_{abcd} U^i_{abcd}
\hat{c}^\dagger_{i a \sigma} \hat{c}^\dagger_{i c \sigma'}
\hat{c}^{\phantom{\dagger}}_{i b \sigma}
\hat{c}^{\phantom{\dagger}}_{i d \sigma'},
\label{eqn.ManyBodyH}
\end{equation}
\end{widetext}
\noindent is formulated in the basis of the Wannier functions~\cite{WannierRevModPhys}, where $\hat{c}^\dagger_{i a \sigma}$ ($\hat{c}_{i a \sigma}$) is the operator of creation (annihilation) of an electron at the orbital $a = xy$, $yz$, $3z^2$$-$$r^2$, $zx$, or $x^2$$-$$y^2$ of the Fe site $i$ with the spin $\sigma= \uparrow$ or $\downarrow$~\cite{footnote1}. The Wannier functions are constructed using the projector-operator technique and the orthonormal LMTO's as the trial functions~\cite{review2008}.

\par The one-electron part of the model Hamiltonian, $\hat{t} = [ t_{ij}^{ab} ]$, is given by the matrix elements of the Kohn-Sham LDA Hamiltonian in the Wannier basis. Since the latter is complete in the subspace of the Fe $3d$ bands, the obtained $\hat{t}$ perfectly reproduces the original LDA bands in this region (Fig.~\ref{fig.elstr})~\cite{review2008}. Then, the matrix elements of $\hat{t}$ with $i \ne j$ stand for the transfer integrals, while the ones with $i=j$ describe the crystal-field effects.

\par
\begin{figure}[b]
\begin{center}
\includegraphics[width=0.4\textwidth]{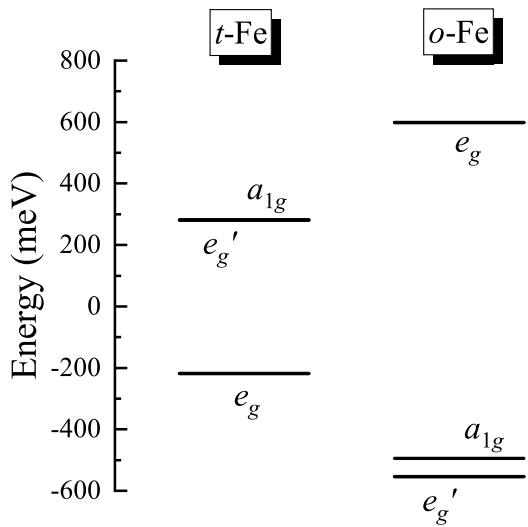}
\end{center}
\caption{
Atomic level splitting at the tetrahedral (left) and octahedral (right) Fe sites.}
\label{fig.CF}
\end{figure}

 The scheme of atomic level splitting (the eigenvalues of $[ t_{ij}^{ab} ]$ for $i=j$) is shown in Fig.~\ref{fig.CF}.
\noindent
\noindent As expected, the $3d$ levels are split into the triply-degenerate $t_{2g} \equiv a_{1g} \oplus e_{g}'$ and doubly-degenerate $e_{g}$ states. In the tetrahedral environment, the $e_{g}$ states are located lower in energy, while in the octahedral one the order of the $t_{2g}$ and $e_g$ levels is reversed. The $t_{2g}$-$e_g$ splitting ($10Dq$) is about $-499$ and $1133$ meV at the $t$-Fe and $o$-Fe sites, respectively, which is reasonable agreement with the results of the Wien2k calculations ($-625$ and $1160$ meV, respectively). The splitting is substantially larger at the $o$-Fe sites, which is consistent with the form of LDA density of states in Fig.~\ref{fig.elstr}, where the $t$-Fe $3d$ states are located near the Fermi level and sandwiched by the $o$-Fe $3d$ states from below and above. In the hexagonal P6$_3$mc symmetry, the $t_{2g}$ levels are further split into non-degenerate $a_{1g}$ and doubly-degenerate $e_{g}'$ states by about $1$ and $59$ meV at the $t$-Fe and $o$-Fe sites, respectively (where the $e_{g}'$ states are located lower in energy). The Wien2k method provides somewhat different scheme of the $t_{2g}$ level splitting: $-126$ and $-53$ meV at the $t$-Fe and $o$-Fe sites, respectively, where the lower energy level is of the $a_{1g}$ symmetry. The difference is related to the asphericity of the Kohn-Sham potential in the Wien2k method. Nevertheless, some portion of this asphericity (and, therefore, the crystal-field splitting) should be subtracted in order to avoid the double-counting problem in the process of solution of the Hubbard model (\ref{eqn.ManyBodyH}), which also includes the nonspherical effects, of the same origin, driven by the screened on-site Coulomb interaction $U^i_{abcd}$~\cite{review2008}. Fortunately, the $t_{2g}$ level splitting is not particularly large and does not affect our finite results: in numerical calculations we used two schemes of the level splitting, obtained in LMTO and Wien2k, and both of them yielded similar conclusion regarding the form of the orbital ordering and interatomic exchange interactions.

\par Thus, from the viewpoint of symmetry and atomic level splitting, one can expect the following scenaria. First of all, the majority-spin states of $t$-Fe and $o$-fe will be fully occupied. Then, 2 minority-spin electrons can reside at the low-lying $e_g$ orbitals of $t$-Fe, resulting in the charge-disproportionated solution $d_{t}^{7}d_{o}^{5}$, which respects the P6$_3$mc symmetry of Fe$_2$Mo$_3$O$_8$. It may look at odds with the scheme of crystal-field splitting (Fig.~\ref{fig.CF}), where the $t_{2g}$ orbitals of $o$-Fe are located lower in energy and therefore are expected to be occupied first. However, we will see in a moment that the $d_{t}^{7}d_{o}^{5}$ solution is also supported by the form of the screened on-site Coulomb interactions. In the case of homogeneous solution $d_{t}^{6}d_{o}^{6}$ (the second scenario), each of the minority-spin electrons at the $t$-Fe and $o$-Fe sites will reside at the degenerate $e_{g}$ and $t_{2g}$ orbitals, respectively, so that the system will tend to lift the degeneracy through the Jahn-Teller distortion and/or orbital ordering.

\par The parameters of screened on-site Coulomb interactions, $\hat{U} = [U^i_{abcd}]$, were calculated using simplified version of the constrained random-phase approximation (RPA)~\cite{Ferdi04}, as explained in Ref.~\onlinecite{review2008}. Each $5$$\times$$5$$\times$$5$$\times5$ matrix $\hat{U} = [U^i_{abcd}]$ can be fitted in terms of the Coulomb repulsion $U=F^0$, the intra-atomic exchange interaction $J = (F^2$$+$$F^4)/14$, and the nonsphericity $B = (9F^2$$-$$5F^4)/441$, where $F^0$, $F^2$, and $F^4$ are the screened radial Slater's integrals~\cite{JPSJ}. The results of such fitting are shown in Table~\ref{tab:scrint}.
\noindent
\begin{table}[t]
\caption{Parameters of screened Coulomb interaction ($U$), exchange interaction ($J$) and nonsphericity ($B$)
for the tetrahedral and octahedral Fr sites in Fe$_2$Mo$_3$O$_8$ (in eV).}
\label{tab:scrint}
\begin{ruledtabular}
\begin{tabular}{lcc}
    & $t$-Fe & $o$-Fe  \\
\hline
$U$ & $1.52$ & $1.80$  \\
$J$ & $0.80$ & $0.78$  \\
$B$ & $0.08$ & $0.07$
\end{tabular}
\end{ruledtabular}
\end{table}
\noindent One can see that the screened $U$ is relatively small. This is understandable considering the electronic structure of Fe$_2$Mo$_3$O$_8$: the Fe $3d$ bands are sandwiched by the Mo $4d$ ones (Fig.~\ref{fig.elstr}), which also have a large weight of the Fe $3d$ states and, therefore, very efficiently screen the Coulomb interactions in the target Fe $3d$ bands~\cite{review2008}. Furthermore, the Coulomb $U$ is smaller at the tetrahedral sites. This is also closely related to the electronic structure of Fe$_2$Mo$_3$O$_8$, where the $t$-Fe $3d$ bands are mainly located near the Fermi level, inside the $o$-Fe ones: since the screening in RPA is governed by the electronic excitations between occupied and unoccupied states, the strongest effect is expected for those states, which are located near the Fermi level. The change of the Coulomb repulsion parameter between tetrahedral and orthorhombic sites, $\Delta U = U^{o} - U^{t}$, is about $0.3$ eV, which does not seem to be large. Nevertheless, it corresponds to the change of the Coulomb potential $\delta v_{\rm C} = \Delta U (n-1) \sim 1.5$ eV for $n=6$, which tends to drive the system into the charge disproportionation regime and formation of the electronic state $d_{t}^{7}d_{o}^{5}$ instead of the charge homogeneous one $d_{t}^{6}d_{o}^{6}$.

\subsection{\label{sec:solution} Solution of the model}
\par
\begin{figure}[t]
\begin{center}
\includegraphics[width=0.47\textwidth]{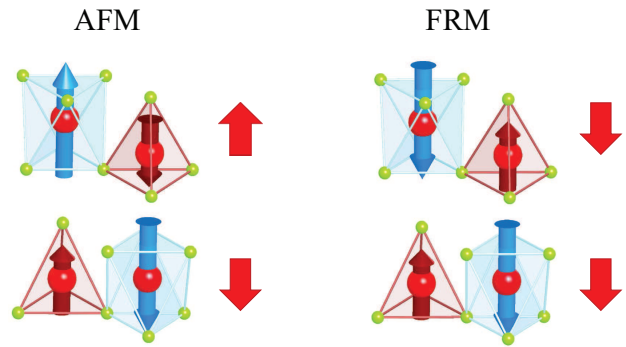}
\end{center}
\caption{
Antiferromagnetic (AFM) and ferrimagnetic (FRM) structure of Fe$_2$Mo$_3$O$_8$. Fe and O atoms are denoted by large and small spheres respectively. The directions of local magnetic moments at the tetrahedral and octahedral sites are shown by small (brown) and big (blue) arrows, respectively. The direction of net magnetization in each layer is shown by fat (red) arrow in from of this layer.}
\label{fig.mstr}
\end{figure}

\begin{figure}[b]
\begin{center}
\includegraphics[width=0.47\textwidth]{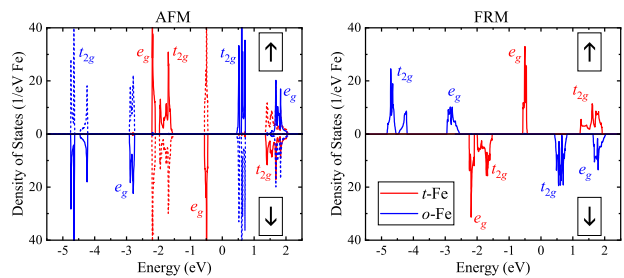}
\end{center}
\caption{
Partial densities of states as obtained in the mean-field Hartree-Fock calculations for the antiferromagnetic (AFM) and ferrimagnetic (FRM) charge disproportionated $d_{t}^{7}d_{o}^{5}$ phases. The contributions of the $t$-Fe and $o$-Fe atoms are shown by red and blue colors, respectively. In the AFM case, the contributions of atoms located in the antiferromagnetically coupled adjacent layers are shown by solid and dashed lines. The Fermi level, defined as the midpoint of the band gap, is at zero energy.}
\label{fig.dos}
\end{figure}

 The model (\ref{eqn.ManyBodyH}) was solved in the mean-field Hartree-Fock (HF) approximation~\cite{review2008} for the AFM and FRM phases (see Fig.~\ref{fig.mstr}) as well as other magnetic configurations, which were used for the construction of the spin model~\cite{SM}.
\noindent
\noindent The straightforward solution of the model (\ref{eqn.ManyBodyH}) leads to the $d_{t}^{7}d_{o}^{5}$ configuration, which is supported by the crystal-field splitting of the atomic $3d$ levels and the values of the Coulomb repulsion $U$ at the $t$-Fe and $o$-Fe sites. The corresponding densities of states are shown in Fig.~\ref{fig.dos}.
\noindent
\noindent As expected, this solution is insulating: the band gap is about $1$ eV and formed between $e_{g}$ states of $t$-Fe and $t_{2g}$ states of $o$-Fe.

\par Nevertheless, we do not rule out the possibility that the obtained charge-disproportionated solution $d_{t}^{7}d_{o}^{5}$ may also be an artifact of calculations, because our model~(\ref{eqn.ManyBodyH}) does not include the double-counting term~\cite{LDAU}. The double-counting term typically serve to subtract the portion of Coulomb and exchange-correlation interactions, which are already included at the level of LDA/GGA (the generalized gradient approximation)~\cite{LDAU}. In the homogeneous case with one type of correlated ions, this correction is reduced to the constant energy shift and, therefore, can be neglected, since calculating the Fermi level we restore status quo. However, if the screened Coulomb repulsion is different at different atomic sites, as in the case of the $t$-Fe and $o$-Fe, such correction can be important.

\par
\begin{figure}[t]
\begin{center}
\includegraphics[width=0.47\textwidth]{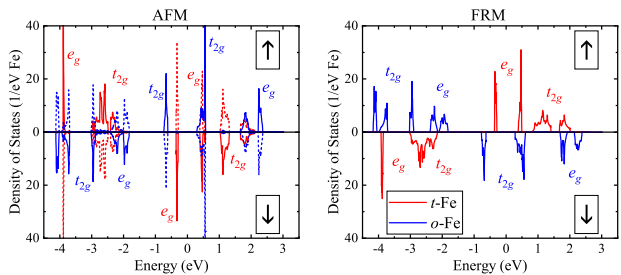}
\end{center}
\caption{
Partial densities of states as obtained in the mean-field Hartree-Fock calculations for the antiferromagnetic (AFM) and ferrimagnetic (FRM) charge homogeneous $d_{t}^{6}d_{o}^{6}$ phases. The contributions of the $t$-Fe and $o$-Fe atoms are shown by red and blue colors, respectively. In the AFM case, the contributions of atoms located in the antiferromagnetically coupled adjacent layers are shown by solid and dashed lines. The Fermi level, defined as the midpoint of the band gap, is at zero energy.}
\label{fig.dosc}
\end{figure}

 Therefore, we have also considered the homogeneous solution $d_{t}^{6}d_{o}^{6}$, which can be obtained in constraint calculations fixing the number of $3d$ electrons at the $t$-Fe and $o$-Fe sites. In fact, the original LDA calculations, where no sizable charge disproportionation have been detected (Fig.~\ref{fig.elstr}), also speak in favor of such homogeneous solution. The corresponding densities of states for the AFM and FRM phases are shown in Fig.~\ref{fig.dosc}.
\noindent
\noindent In this case, the on-site Coulomb interactions lift the orbital degeneracy of the $t$-Fe $e_{g}$ and $o$-Fe $t_{2g}$ levels through the formation of the orbitally ordered state, which breaks the P6$_3$mc symmetry, opens the bang gap of about $0.5$ eV, and minimizes the energy of interatomic exchange interactions~\cite{KugelKhomskii}.

\par
\begin{figure}[b]
\begin{center}
\includegraphics[width=4cm]{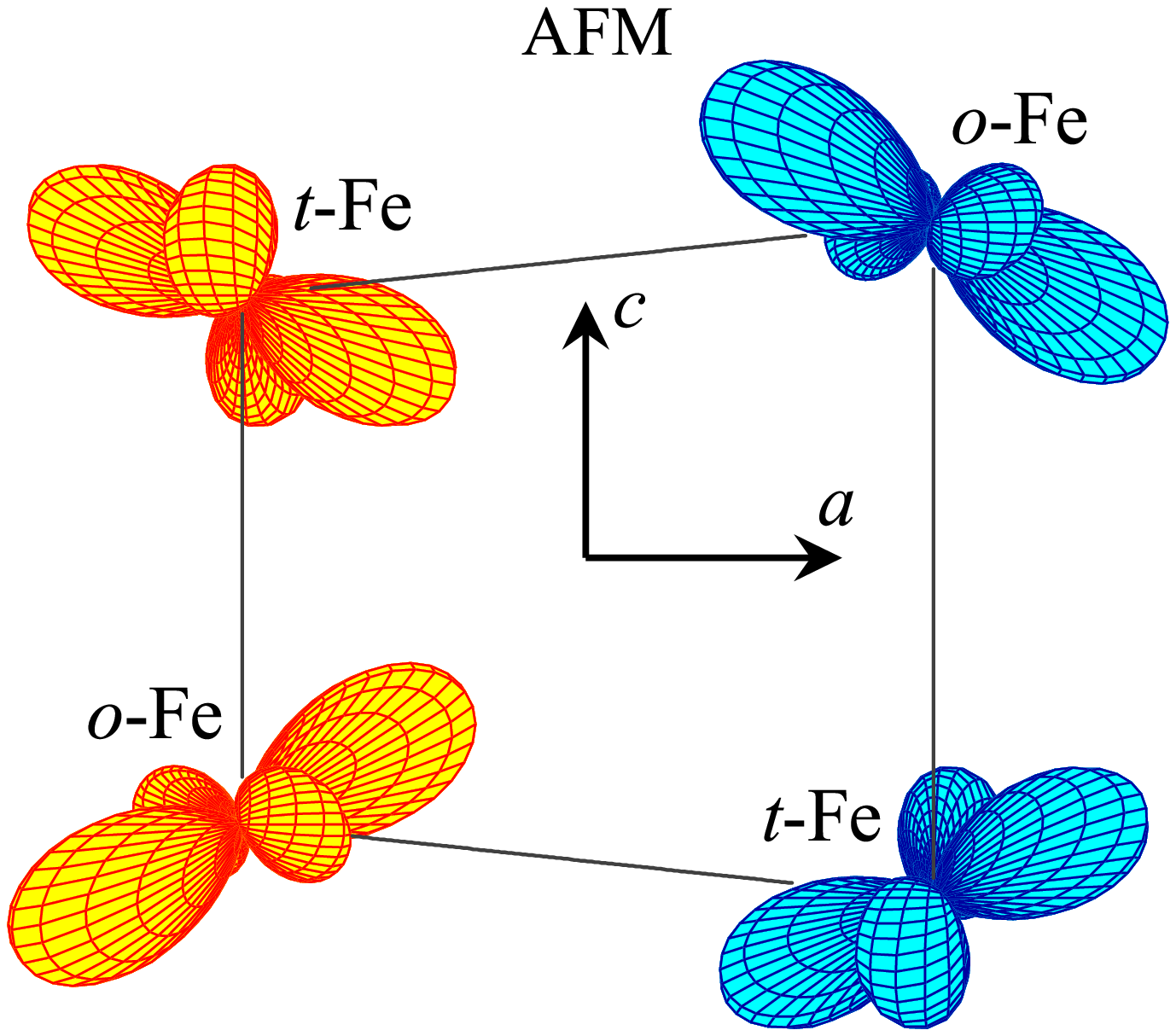}
\includegraphics[width=4cm]{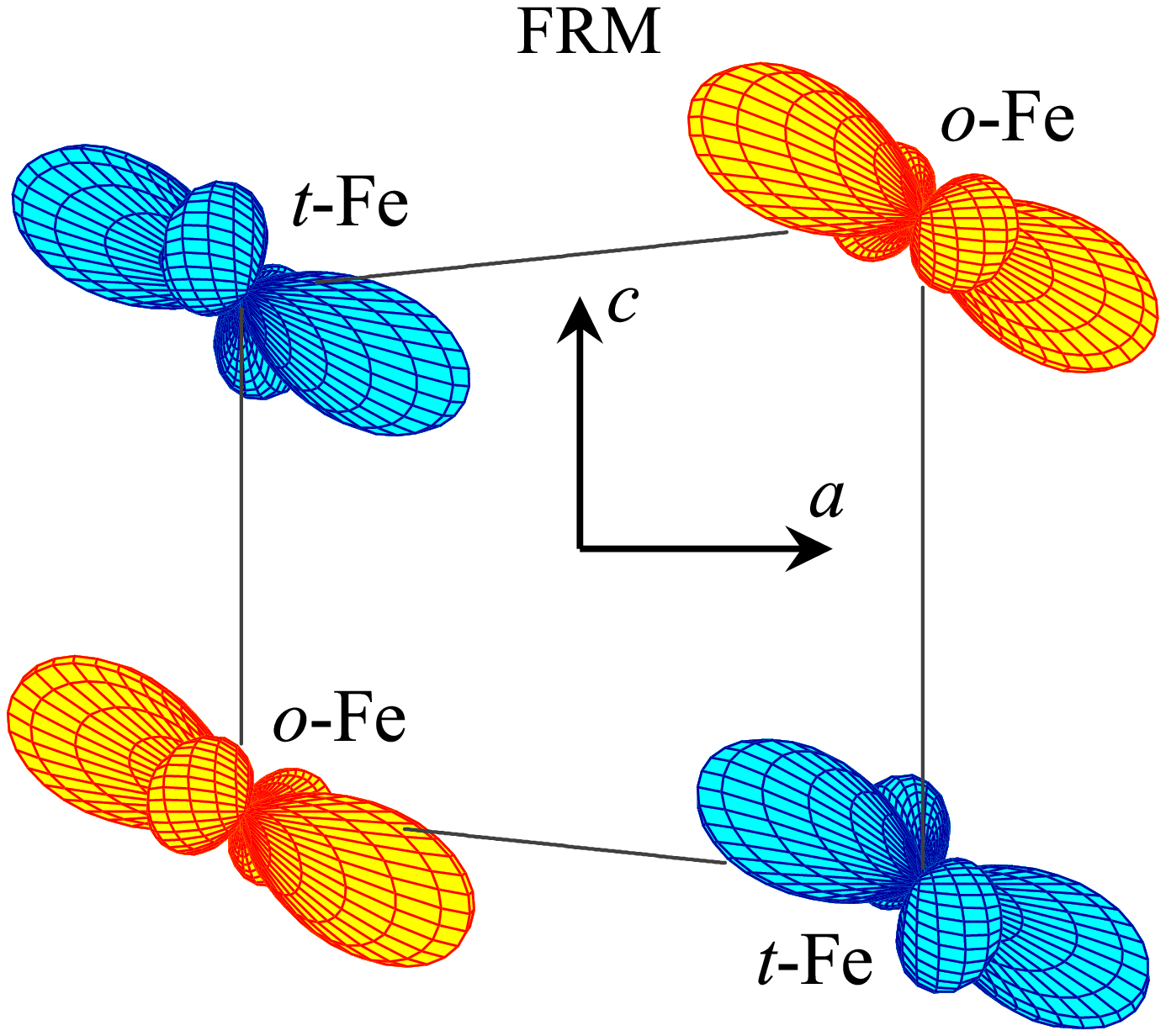}
\end{center}
\caption{
Orbital ordering obtained in constrained Hartree-Fock calculations for the configuration $d_{t}^{6}d_{o}^{6}$ in the case of the antiferromagnetic (AFM) and ferrimagnetic (FRM) spin order. A single occupied orbital of minority spin is shown. }
\label{fig.OO}
\end{figure}

 In order to visualise this orbital ordering, we plot the density formed by one minority-spin electron around each Fe site, which was obtained by integrating the states in the energy window $[-1,0]$ eV in Fig.~\ref{fig.dosc}. The results are shown in Fig.~\ref{fig.OO} for the AFM and FRM phases.
\noindent
\noindent As expected, the change of the spin order from AFM to FRM leads to the change of the orbital order and the spacial reorientation of the occupied minority-spin orbitals so to further stabilize the given spin order~\cite{KugelKhomskii}. Loosely speaking, the AFM coupling between nearest-neighbor sites along the $\boldsymbol{c}$ axis, realized in the FRM phase, coexists with the ``ferro'' orbital order, where the occupied minority-spin orbitals in the bond are oriented in a similar way. On the contrary, the ferromagnetic coupling along $\boldsymbol{c}$ in the AFM phase coexists with the ``antiferro'' orbital order, where the occupied orbitals form some angle with respect to each other. In other words, in the FRM case the system tends to fill the same orbitals for $o$-Fe and $t$-Fe along $\boldsymbol{c}$ in order to minimize the energy of superexchange interactions between these and other orbitals, which have considerable overlap.

\par Finally, we note that the FRM $d_{t}^{6}d_{o}^{6}$ solution corresponds to the compensated ferrimagnetic case, where the $t$-Fe and $o$-Fe sublattices are inequivalent, but the net spin magnetic moment is equal to zero.

\subsection{\label{sec:J} Interatomic exchange interactions}
\par The interatomic exchange interactions can be evaluated by mapping the total energy change caused by the reorientation of spins onto the Heisenberg model~\cite{JHeisenberg}:
\noindent
\begin{equation}
{\cal H}_S = -\frac{1}{2} \sum_{ij} J_{ij} \boldsymbol{e}_{i} \cdot \boldsymbol{e}_{j},
\label{eqn:Heisenberg}
\end{equation}
\noindent where $\boldsymbol{e}_{i}$ is the \emph{direction} of spin at the site $i$. In order to evaluate $J_{ij}$, we used two different techniques. The first one is based on finite rotations of spins, where $J_{ij}$ is related to the total energies of several collinear magnetic configurations obtained by aligning each of the four Fe spins in the unit cell either up or down. The method is standard and widely used in electronic structure community for the analysis of the magnetic properties.

\par The second method is based on the infinitesimal rotations of spins near the equilibrium, where $J_{ij}$ are obtained in the second order perturbation theory with respect to the rotations of the self-consistent HF potentials at the sites $i$ and $j$~\cite{JHeisenberg,review2008}:
\noindent
\begin{equation}
J_{ij} = \frac{1}{2\pi} {\rm Im} \int_{- \infty}^{\varepsilon_{\rm F}} d \varepsilon \, {\rm Tr}_L \left\{
\Delta \hat{V}_{i} \hat{G}_{ij}^{\uparrow}(\varepsilon)
\Delta \hat{V}_{j} \hat{G}_{ji}^{\downarrow}(\varepsilon)
\right\}.
\label{eqn:Jij}
\end{equation}
\noindent Here, $\hat{G}^{\uparrow, \downarrow}(\varepsilon)$ is the one-electron Green's for the majority and minority spin states, $\Delta \hat{V}_{i} = \hat{V}_{i}^\uparrow - \hat{V}_{i}^\downarrow$ is the spin part of the HF potential at the site $i$, $\varepsilon_{\rm F}$ is the Fermi energy, and ${\rm Tr}_L$ denotes the trace over the orbital indices. Generally, the parameters $J_{ij}$ calculated using the second technique depend on the magnetic state, thus reflecting the change of the electronic structure and the orbital ordering. The comparison of such parameters, calculated in different magnetic states, presents a test for the validity of the Heisenberg model, which can be defined locally, for the infinitesimal spin rotations, but not necessary globally, to describe the energies of all possible spin configurations where each spin can have an arbitrary direction, irrespectively on the direction of its neighboring spins.

\par
\begin{figure}[t]
\begin{center}
\includegraphics[width=0.47\textwidth]{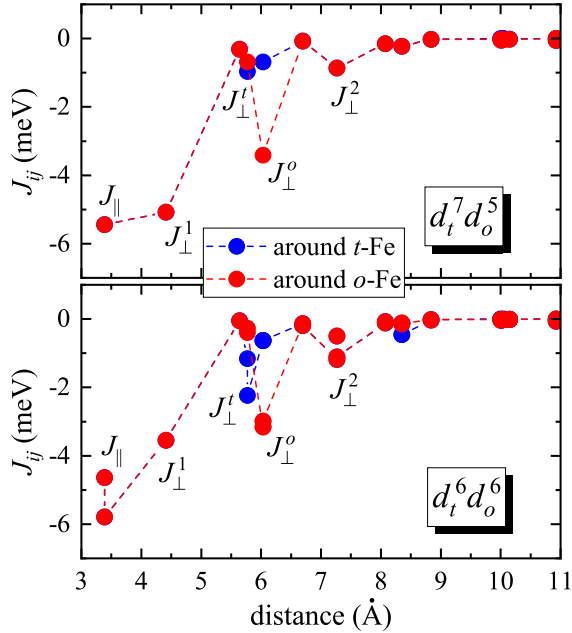}
\end{center}
\caption{
Distance-dependence of exchange interactions around the tetrahedral and octahedral Fe sites as obtained in the Green's function method for the antiferromagnetic $d_{t}^{5}d_{o}^{7}$ and $d_{t}^{6}d_{o}^{6}$ solutions. Main exchange interactions are labeled and explained in Fig.~\ref{fig.str}.}
\label{fig.J}
\end{figure}

 The results of Green's function calculations are summarized in Fig.~\ref{fig.J} and the main exchange interactions are explained in Fig.~\ref{fig.str}.
\noindent
\noindent Somewhat surprisingly, the exchange interactions exhibit quite similar behavior for the solutions $d_{t}^{7}d_{o}^{5}$ and $d_{t}^{6}d_{o}^{6}$.  Furthermore, we note the following: (i) The orbital ordering accompanying the $d_{t}^{6}d_{o}^{6}$ solution for the AFM and FRM states lowers the P6$_3$mc symmetry. Such symmetry lowering is manifested in somewhat different values of the exchange parameters, which are realized in the crystallographically equivalent bonds, as is clearly seen for $J_{\parallel}$, $J_{\perp}^{t}$ and $J_{\perp}^{2}$ in the lower panel of Fig.~\ref{fig.J}. Nevertheless, this difference is not particularly large (for instance, in comparison with the difference between $J_{\parallel}$, $J_{\perp}^{1}$, and other interactions). Therefore, in the first approximation one can average the exchange parameters over the crystallographically equivalent bonds and neglect the difference between them. Such problem does not occurs for the solution $d_{t}^{7}d_{o}^{5}$, which respects the P6$_3$mc symmetry; (ii) Apart from the symmetry lowering, which can be different for the AFM and FRM states reflecting the difference in the orbital ordering, the averaged parameters reveal very similar behavior for the AFM and FRM states~\cite{SM}; (iii) Very similar set of exchange parameters can be obtained by mapping the energies of collinear magnetic configurations and flipping each spin instead of rotating it by an infinitesimal angle (Table~\ref{tab:modelJ}).
\noindent
\begin{table*}[t]
\caption{Parameters of exchange interactions (in meV) obtained by mapping the total energies for the charge disproportionate ($d_{t}^{7}d_{o}^{5}$) and homogeneous ($d_{t}^{6}d_{o}^{6}$) solutions of the effective electron model onto the isotropic spin model. The corresponding averaged parameters obtained by using Green's function perturbation theory technique for the infinitesimal spin rotations are given in parentheses.}
\label{tab:modelJ}
\begin{ruledtabular}
\begin{tabular}{lcccc}
                     & $J_{\parallel}$   & $J_{\perp}$         & $J_{\perp}^{o}$   & $J_{\perp}^{t}$   \\
\hline
$d_{t}^{7}d_{o}^{5}$ & $-5.63$ ($-5.44$) & $-11.90$ ($-10.23$) & $-3.77$ ($-3.41$) & $-0.98$ ($-0.68$) \\
$d_{t}^{6}d_{o}^{6}$ & $-5.74$ ($-5.40$) & $-10.26$ ($-9.13$)  & $-3.20$ ($-3.04$) & $-0.90$ ($-0.63$)
\end{tabular}
\end{ruledtabular}
\end{table*}
\noindent These arguments suggest that the spin model (\ref{eqn:Heisenberg}) is well defined and can be used for the analysis magnetic properties of Fe$_2$Mo$_3$O$_8$ in the wide temperature range.

\par All $J_{ij}$ are antiferromagnetic. The AFM coupling between $t$-Fe and $o$-Fe in each layer is stabilized by $J_{\parallel}$, which is the strongest interaction in the system. The magnetic ordering between the layers results from the competition of three main interactions: the nearest-neighbor (nn) interaction $J_{\perp}^1$ between $t$-Fe and $o$-Fe, together with $J_{\parallel}$, tends to stabilize the FRM phase, while the next-nn interactions $J_{\perp}^{t}$ and $J_{\perp}^{o}$ operating, respectively, in the sublattices $t$-Fe and $o$-Fe favor (again, together with $J_{\parallel}$) the AFM alignment. Furthermore, the effect of $J_{\perp}^{1}$ is strengthened by 2nd neighbor interactions $J_{\perp}^{2}$ between $t$-Fe and $o$-Fe: although $J_{\perp}^{2}$ is considerably smaller, the number of such bonds is large (see Fig.~\ref{fig.str}), making the total contribution comparable with $J_{\perp}^{1}$. Thus, the relevant parameter responsible for the emergence of the FRM order is $J_{\perp} = J_{\perp}^{1}+6J_{\perp}^{2}$. Considering the numbers of bonds, one can find the following condition for the stability of the AFM phase relative to the FRM one: $|J_{\perp}| < 3 |J_{\perp}^{t} + J_{\perp}^{o}|$, which is satisfied for both $d_{t}^{7}d_{o}^{5}$ and $d_{t}^{6}d_{o}^{6}$. Nevertheless, the AFM structure is not the ground state of the model: the competition of $J_{\parallel}$, $J_{\perp}$, and $J_{\perp}^{o}$ ($J_{\perp}^{t}$) should lead to the noncollinear magnetic order with the propagation vector close to $\boldsymbol{q} = (0,0,1/2)$~\cite{SM}. It would be interesting to chesk this point experimentally. Finally, the exchange interaction $J_{\perp}^{t}$ is considerably weaker than $J_{\perp}^{o}$, which has important consequences on the magnetic properties of Fe$_2$Mo$_3$O$_8$: with the increase of the temperature ($T$), the magnetization in the $t$-Fe sublattice will tend to vanish faster than in the $o$-Fe one (which is quite expected for the systems with different magnetic sublattices~\cite{CoV2O4}). Therefore, even for the homogeneous solution $d_{t}^{6}d_{o}^{6}$, where the net magnetization is zero at $T=0$, both in the AFM and FRM case, one can expect appearance of finite net magnetization at finite $T$, which couples to the magnetic field and can be used for the switching between the AFM and FRM phases.

\subsection{\label{sec:P} Parameters of electric polarization}
\par We assume that the magnetic part of the electric polarization parallel to the $z$ axis can be described by the following expression:
\noindent
\begin{equation}
P^{z} = \frac{1}{2} \sum_{ij} P_{ij} \boldsymbol{e}_{i} \cdot \boldsymbol{e}_{j},
\label{eqn:Pz}
\end{equation}
which is similar to Eq.~(\ref{eqn:Heisenberg}) for the exchange interaction energy. In principle, Eq.~(\ref{eqn:Pz}) can be derived rigorously, by applying the Berry-phase theory of electric polarization~\cite{FE_theory} to the model (\ref{eqn.ManyBodyH})~\cite{PRB2012} and considering the limit of large $U$, as is typically done in the theories of double exchange and superexchange interactions for the spin Hamiltonian~(\ref{eqn:Heisenberg}) without spin-orbit coupling~\cite{PRB2014,PRB2019}. Nevertheless, since interatomic exchange interactions $J_{ij}$ are well reproduced by mapping the total energies obtained in the self-consistent Hartree-Fock calculations for a limited number of magnetic configuration, we employ here a similar strategy for $P^{z}$ and derive the parameters $P_{ij}$ by mapping the values of electric polarization obtained in the same calculations onto Eq.~(\ref{eqn:Pz}) and assuming that, similar to $J_{ij}$, the main details of $P^{z}$ can be described by four independent parameters: $J_{\parallel}$, $J_{\perp}$, $J_{\perp}^{o}$, and $J_{\perp}^{t}$. They are listed in Table~\ref{tab:modelP}.
\noindent
\begin{table}[b]
\caption{Parameters of electric polarization (in $\mu$C/m$^2$) obtained by mapping the polarizations obtained for charge disproportionate ($d_{t}^{7}d_{o}^{5}$) and homogeneous ($d_{t}^{6}d_{o}^{6}$) solutions of the effective electron model onto the isotropic spin model.}
\label{tab:modelP}
\begin{ruledtabular}
\begin{tabular}{lcccc}
                     & $P_{\parallel}$  & $P_{\perp}$       & $P_{\perp}^{o}$ & $P_{\perp}^{t}$  \\
\hline
$d_{t}^{7}d_{o}^{5}$ & $-384$           & $-122$            & $-302$          & $194$            \\
$d_{t}^{6}d_{o}^{6}$ & $\phantom{-3}41$ & $\phantom{-1}24$  & $-194$          & $\phantom{1}66$
\end{tabular}
\end{ruledtabular}
\end{table}
\noindent Unlike $J_{ij}$, the parameters $P_{ij}$ differ substantially in the case of $d_{t}^{7}d_{o}^{5}$  and $d_{t}^{6}d_{o}^{6}$. In the former case, all parameters are large and equally important, while in the latter case $P_{\perp}^{o}$ clearly prevails. Somewhat unexpectedly, we have found large $P_{\parallel}$ for charge disproportionated configuration $d_{t}^{7}d_{o}^{5}$. Indeed, $P^{z}$ is proportional to $\Delta z$ (the difference of atomic $z$-coordinates in the bond), which is rather small for the nn in-plane bonds (about $0.6$~\AA). On the other hand, the ionic charge difference between $t$-Fe$^{1+}$ and $o$-Fe$^{3+}$ is large, which readily compensates the smallness of $\Delta z$. In the charge neutral regime, $d_{t}^{6}d_{o}^{6}$, $P_{\parallel}$ is expectedly small (and is fully associated with the redistribution of the tails of the Wannier functions at the $t$-Fe and $o$-Fe sites~\cite{PRB2014,PRB2019}).

\par In principle, the model can be further extended to include antisymmetric and anisotropic effects driven by the relativistic spin-orbit coupling. The corresponding expressions can be found in Ref.~\cite{PRB2019}. However, since the magnetic transition takes place between two collinear configurations, AFM and FRM, it is reasonable to expect that the main contribution to the change of $P^{z}$ is isotropic and described by Eq.~(\ref{eqn:Pz}).

\section{\label{sec:discussion} Discussion}
\par
\begin{figure}[t]
\begin{center}
\includegraphics[width=0.47\textwidth]{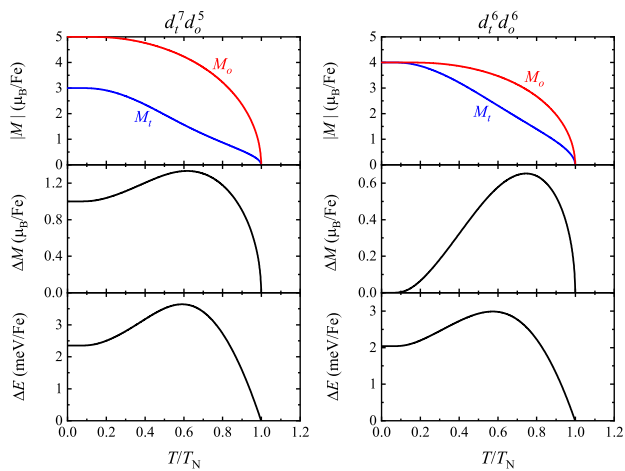}
\end{center}
\caption{
Results of molecular-field theory for the spin model: temperature dependence of magnetization, $M$, at the $t$-Fe and $o$-Fe sites, net magnetic moment in the honeycomb layer, $\Delta M = (|M_{o}|-|M_{t}|)/2$, recalculated per one Fe site, and the total energy difference, $\Delta E$, between ferrimagnetic and antiferromagnetic phases calculated using parameters for the $d_{t}^{7}d_{o}^{5}$ and $d_{t}^{6}d_{o}^{6}$ states.}
\label{fig.MFE}
\end{figure}

\begin{figure}[t]
\begin{center}
\includegraphics[width=0.47\textwidth]{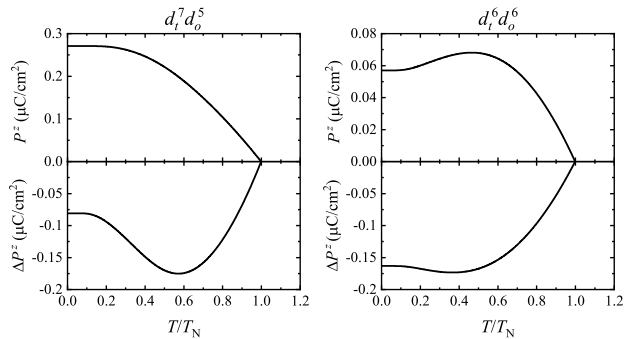}
\end{center}
\caption{
Results of molecular-field theory for the spin model: temperature dependence of the spin-dependent part of the electric polarization ($P^{z}$) in the antiferromagnetic phase and the polarization jump ($\Delta P^{z}$) caused by the antiferromagnetic-to-ferrimagnetic transition calculated using parameters for the $d_{t}^{7}d_{o}^{5}$ and $d_{t}^{6}d_{o}^{6}$ states.}
\label{fig.MFP}
\end{figure}

 Much insight can be gained from the solution of the spin model (\ref{eqn:Heisenberg}) in the molecular-field approximation (MFA). Namely, the molecular field corresponding to the spin Hamiltonian (\ref{eqn:Heisenberg}) is given by
\noindent
\begin{equation}
h_{i} = - \sum_j J_{ij} m_{j} (T)
\label{eqn:MF}
\end{equation}
\noindent where $m_{j}(T) = M_{j}(T)/| M_{j}(0) |$ is the relative magnetization at the site $j$. Then, $m_{i} (T)$ can be found from the temperature average $M_{j} = 2 \langle \hat{S}_{j}^{z} \rangle$ of the spin operator $\hat{S}_{j}^{z}$ in the molecular field $h_{i}$:
\noindent
\begin{equation}
m_{i} (T) = \frac{h_{i}}{|h_{i}|} B_{S_{i}} \left( \frac{|h_{i}|}{k_BT} \right),
\label{eqn:SMeanField}
\end{equation}
\noindent where $B_{S_{i}}$ is the Brillouin function for the spin $S_{i}$~\cite{Mattis}. The equations (\ref{eqn:MF}) and (\ref{eqn:SMeanField}) are solved self-consistently and the N\'eel temperature ($T_{\rm N}$) is defined as the minimal temperature for which $m_{i} (T) = 0$. Then, the spin-dependent part of the polarization in the AFM state, the total energy difference between the FRM and AFM phases, and the polarization jump caused by the AMF-to-FRM transition can be evaluated as
\noindent
\begin{equation}
P^{z} = \left( 2P_{\perp} - 6P_{\parallel} \right)|m_{t}||m_{o}| -6P_{\perp}^{t}m_{t}^{2} -6P_{\perp}^{o}m_{o}^{2},
\label{eqn:MFP}
\end{equation}
\noindent
\begin{equation}
\Delta E = 4J_{\perp}|m_{t}||m_{o}| -12J_{\perp}^{t}m_{t}^{2} -12J_{\perp}^{o}m_{o}^{2},
\label{eqn:MFdE}
\end{equation}
\noindent and
\noindent
\begin{equation}
\Delta P^{z} = -4P_{\perp}|m_{t}||m_{o}| +12P_{\perp}^{t}m_{t}^{2} +12P_{\perp}^{o}m_{o}^{2},
\label{eqn:MFdP}
\end{equation}
\noindent respectively. Unless specified otherwise, we use the parameters listed in Tables~\ref{tab:modelJ} and \ref{tab:modelP}. The results are summarized in Figs.~\ref{fig.MFE} and~\ref{fig.MFP}.
\noindent
\noindent

\par The molecular field estimate for $T_{\rm N}$ is about $132$ and $128$ K for $d_{t}^{7}d_{o}^{5}$ and $d_{t}^{6}d_{o}^{6}$, respectively. Quite expectedly, similar sets of parameters $J_{ij}$ (see Table~\ref{tab:modelJ}) yield similar values of $T_{\rm N}$. Thus, from this point of view the solutions $d_{t}^{7}d_{o}^{5}$ and $d_{t}^{6}d_{o}^{6}$ are ``indistinguishable''. More rigorous estimate for $T_{\rm N}$ can be obtained by considering Tyablikov's RPA~\cite{tyab}, generalized to the case of multiple magnetic sublattices~\cite{TCRPA} and noncollinear magnetic ground state~\cite{SM}, which is expected in both $d_{t}^{7}d_{o}^{5}$ and $d_{t}^{6}d_{o}^{6}$ models for Fe$_2$Mo$_3$O$_8$ (see Sec.~\ref{sec:J}). The RPA yields $T_{\rm N} = 55$ and $54$ K for $d_{t}^{7}d_{o}^{5}$ and $d_{t}^{6}d_{o}^{6}$, respectively. The latter estimates are close to the experimental $T_{\rm N} = 60$ K~\cite{Wang_SciR,Kurumaji_PRX}, while the MFA values are typically overestimated. The large difference between the MFA and RPA is related to the existence of weakly dispersive regions of magnon energies, which are nearly degenerate with the ground state~\cite{SM}. We have also used the full set of parameters, obtained in the Green's function calculations for $d_{t}^{7}d_{o}^{5}$ (Fig.~\ref{fig.J}), which obeys the crystallographic P6$_3$mc symmetry. This yields slightly smaller value of $T_{\rm N} = 105$ and $32$ K in MFA and RPA, respectively. Thus, even though the MFA substantially overestimates $T_{\rm N}$, it is still interesting to explore the abilities of this approximation for the description of magnetoelectric properties of Fe$_2$Mo$_3$O$_8$, at least on the semi-quantitative level.

\par Since $|J_{\perp}^{t}| \ll |J_{\perp}^{o}|$, the magnetization in the $t$-Fe and $o$-Fe sublattice exhibits different temperature dependence, where $| M_{t} |$ tends to decrease more rapidly than $| M_{o} |$ with the increase of $T$. Then, the temperature dependence of the net magnetization, $\Delta M = (|M_{o}|-|M_{t}|)/2$, will be nonmonotonous, with some ``optimal value'' corresponding to the maximum of $\Delta M(T)$, for which one can achieve the largest energy gain caused by the interaction with the external magnetic field. This effect is especially important for $d_{t}^{6}d_{o}^{6}$, where the spins in the $t$-Fe and $o$-Fe sublattices exactly cancel each other at $T=0$, thus excluding a linear coupling with the magnetic field. Nevertheless, at finite $T$, such cancellation does not occur, giving rise to the net magnetization in each honeycomb layer, the direction of which can be controlled by the magnetic field so to cause the AFM-FRM transition.

\par The key question is whether the AFM-FRM transition can be induced by experimentally accessible magnetic field, $H_{c}$, which depends on $T$ and varies from about $2$ ${\rm T}$ at $T \sim 0.97~T_{\rm N}$ till $14$ ${\rm T}$ at $T \sim 0.58~T_{\rm N}$~\cite{Wang_SciR,Kurumaji_PRX}. Although theoretical $H_{c}$, which can be estimated as $H_{c} = \frac{\Delta E}{\mu_{\rm B} | \Delta M |}$, shows the same tendency, it is overestimated in comparison with the experiment: for instance, at $T \sim 0.97~T_{\rm N}$ our $H_{c}$ is about $20$ ${\rm T}$ and further increases with the decrease of $T$. One reason may be the overestimation of $\Delta E$ in MFA. Moreover, this $\Delta E$ has a maximum as a function $T$: since $|m_{t}|$ decreases more rapidly, the last term in Eq.~(\ref{eqn:MFdE}) starts to prevail at elevated $T$ and additionally stabilizes AFM order relative to the FRM one. This worsens the agreement with the experimental data for $H_{c}$. Another reason is that we do not consider the lattice effects, assuming that the AFM and FRM phases are described by the same crystal structure, while in reality the lattice relaxation in the FRM phase will certainly decrease the value of $\Delta E$.

\par Thus, from the viewpoint of magnetism, the main difference between the $d_{t}^{7}d_{o}^{5}$ and $d_{t}^{6}d_{o}^{6}$ scenarios is that in the former case $\Delta M$ remains finite even at small $T$, leaving possibility of the AFM-FRM transition in the magnetic field. This could be checked experimentally and according to our estimates it will require $H_{c} \sim 40$ ${\rm T}$.

\par The behavior of spin-dependent part of the electric polarization is sensitive to the charge state of the Fe ions. Since the parameters of polarization are generally smaller for the homogeneous $d_{t}^{6}d_{o}^{6}$ state (see Table~\ref{tab:modelP}), $P^{z}$ is also smaller (by about factor 4 in comparison with $d_{t}^{7}d_{o}^{5}$). The obtained $P^{z}(0) = 0.27$ $\mu$C/cm$^2$ in the $d_{t}^{7}d_{o}^{5}$ model is comparable with the experimental value of about $0.34$ $\mu$C/cm$^2$~\cite{Wang_SciR}. Nevertheless, the overall shape of $P^{z}(T)$ is quite different: the experimental dependence $P^{z}(T)$ exhibits the jump at $T_{\rm N}$, which may signal that the magnetic transition is accompanied by the structural one~\cite{Wang_SciR}, while the theoretical $P^{z}$ decreases steadily down to $T_{\rm N}$.

\par The theoretical $P^{z}$ for $d_{t}^{6}d_{o}^{6}$ has a clear maximum at $T \sim 0.5~T_{\rm N}$, similar to the behavior of $\Delta E$ (Fig.~\ref{fig.MFE}). This is because $P_{\perp}^{o}$ is the strongest parameter in the case of $d_{t}^{6}d_{o}^{6}$ (see Table~\ref{tab:modelP}), which clearly dominates with the increase of $T$ when other contributions to Eq.~(\ref{eqn:MFP}) decrease due to more rapid decrease of $|m_{t}|$. On the other hand, $P^{z}$ in $d_{t}^{7}d_{o}^{5}$ is nearly monotonous function of $T$: in this case, the effect of $P_{\perp}^{o}$ is partly compensated by $P_{\perp}^{t}$, so that the temperature dependence of $P^{z}$ is mainly controlled by strong $P_{\parallel}$ in the first term in Eq.~(\ref{eqn:MFP}). Thus, in principle, the temperature dependence of $P^{z}$ can be used to distinguish experimentally between the configurations $d_{t}^{7}d_{o}^{5}$ and $d_{t}^{6}d_{o}^{6}$.

\par Nevertheless, both scenaria yield a comparable polarization jump $\Delta P^{z}$, caused by the AFM-FRM transition near $T_{\rm N}$ (see Fig.~\ref{fig.MFP}). First, $\Delta P^{z}$ does not depend on $P_{\parallel}$. Then, the effect of strong $P_{\perp}^{o}$ in the case of $d_{t}^{7}d_{o}^{5}$ is compensated by $P_{\perp}$ and $P_{\perp}^{t}$, which are also strong, while in the case of $d_{t}^{6}d_{o}^{6}$, $\Delta P^{z}$ is mainly controlled by $P_{\perp}^{o}$. The value of $\Delta P^{z}$ at $T \sim 0.8~T_{\rm N}$ is about $-0.1$ $\mu$C/cm$^2$, which is comparable with the experimental data~\cite{Wang_SciR,Kurumaji_PRX}. Finally, we note also that $P^{z}$ is positive while $\Delta P^{z}$ is negative, which is also consistent with the experimental situation.

\section{\label{sec:conc}Summary and Conclusions}
\par The magnetic exchange interactions and the origin of giant magnetoelectric effect in Fe$_2$Mo$_3$O$_8$ have been studied on the basis of microscopic toy model derived for the magnetic Fe $3d$ states from the first-principles electronic structure calculations. In spites of its simplicity, the model provides rather rich physics and accounts for the magnetic properties of Fe$_2$Mo$_3$O$_8$ on the semi-quantitative level. Particularly, we propose two scenaria for the magnetic behavior of Fe$_2$Mo$_3$O$_8$. The first one is based on the homogeneous distribution of the Fe$^{2+}$ ions amongst the $t$- and $o$-sites, while the second one involves the charge disproportionation $2$Fe$^{2+}$ $\rightarrow$ Fe$^{1+}$$+$Fe$^{3+}$ with somewhat exotic ionization state $1+$ at the $t$-sites. Both scenaria lead ro similar sets of interatomic exchange interactions, which are consistent with available experimental data and explain the origin of the AFM and FRM phases. The crucial test to distinguish between the $d_{t}^{6}d_{o}^{6}$ and $d_{t}^{7}d_{o}^{5}$ configurations is the net magnetization in the honeycomb layer at low $T$, which is expected to vanish (and emerge only at elevated $T$) in the case of $d_{t}^{6}d_{o}^{6}$, but remains finite in the case of $d_{t}^{7}d_{o}^{5}$ in the molecular-field approximation, thus giving a possibility to control this nagnetization and induce the AFM-FRM transition by applying magnetic field.

\par Our calculations reproduce the order of magnitude of the experimentally observed giant magnetoelectric effect in Fe$_2$Mo$_3$O$_8$, which we attribute to the electronic polarization related to the change of the electronic structure depending on the magnetic state, but for the fixed crystal structure~\cite{Picozzi}. However, the quantitative description of the temperature dependence of the polarization change will probably require the lattice effects, as was suggested in Ref.~\cite{Wang_SciR}.

\par Another interesting problem, which was not addressed in the present work, is the effects of relativistic spin-orbit (SO) interaction and the orbital magnetism, which are expected to play an important role especially in the $d_{t}^{6}d_{o}^{6}$ configuration with the orbital degeneracy. Nevertheless, the problem is rather complex to be systematically studied in the present publication. Briefly, in the case of $d_{t}^{7}d_{o}^{5}$, our mean-field HF calculations for the available experimental P6$_3$mc structure with the SO coupling yield unquenched orbital moment of about $0.4$ $\mu_{\rm B}$ at the $t$-Fe sites, which has the same direction as the spin one, according to third Hund's rule. The orbital moment at the $o$-Fe sites is negligibly small as expected for the $d^{5}$ configuration. Thus, the orbital magnetization contributes to the net magnetic polarization in the honeycomb layer, though this contribution is not particularly strong in comparison with the spin one. In the $d_{t}^{6}d_{o}^{6}$ case, the SO interaction lifts the orbital degeneracy lowering the P6$_3$mc symmetry and resulting in the \textit{canted} spin state. In the ground state, the canting is such that the $z$ ($c$) components of magnetic moments are ordered as AFM, while the $xy$ ($ab$) components form the FRM structure. Beside spin, we also expect the orbital magnetization of the order of $0.6$ $\mu_{\rm B}$ at the $t$-Fe and $o$-Fe sites. Thus, if this scenario is correct, the AFM-FRM transition can be tuned continuously, by applying the magnetic field \textit{in the $xy$ plane} and thus tune the value of the electric polarization.

\section{Acknowledgements}
We are grateful to D.-J. Huang, S.-W. Cheong, Z. Hu, A. Ushakov, S. Nikolaev, and D. Khomskii for valuable discussions. This work was supported by the Russian Science Foundation through RSF 17-12-01207 research grant.


\begin{thebibliography}{99}
\bibitem{footnote1}
All model parameters are available upon request.
\end{thebibliography}
\end{document}